\documentclass[prl,aps,twocolumn,superscriptaddress,showpacs]{revtex4-1}

\usepackage{graphicx}% Include figure files
\usepackage{epstopdf}

\begin{document}

\title{Reply to comment by Mayers \textit{et al.} on ``High energy neutron scattering from hydrogen using a direct geometry spectrometer"}

\author{C. Stock}
\affiliation{NIST Center for Neutron Research, 100 Bureau Drive, Gaithersburg, Maryland 20899, USA}
\affiliation{Indiana University, 2401 Milo B. Sampson Lane, Bloomington, Indiana 47404, USA}

\author{R.A. Cowley}
\affiliation{Oxford Physics, Clarendon Laboratory, Parks Road, Oxford, United Kingdom OX1 3PU, UK}
\affiliation{Diamond, Rutherford Appleton Laboratory, Chilton, Didcot, OX11 0QX, UK}

\date{\today}


\maketitle

	Our paper entitled ``High-energy neutron scattering from hydrogen using a direct geometry spectrometer" (Ref. \onlinecite{Stock10:81}) describes an investigation of the validity of conventional scattering theory on the cross section of hydrogen using a direct geometry spectrometer.  Contrary to previous results using indirect geometry machines, which observe a 20-40\% deficit in the cross section, we find the cross section is constant and therefore consider that the previous results are an experimental artifact from the use of indirect geometry spectrometers.  The comment described in Ref. \onlinecite{Mayers09:xx} provides a detailed discussion regarding the resolution function in the case of direct and indirect geometry neutron scattering instruments at pulsed sources.  Based on this analysis it is claimed that the conclusions made from data with a direct geometry spectrometer (outlined in our publication) are invalid.

	In this reply, we point out several criticisms of the analysis in Ref. \onlinecite{Mayers09:xx} and show that the comment does not change the underlying conclusion presented by ourselves that there is no measurable deficit in the scattering cross section of hydrogen.  We therefore consider that our original conclusions are correct namely that the previous anomalies in the cross section are due to experimental effects related to the use of indirect geometry spectrometers.

\subsection{1) Impulse Approximation}

	Ref. \onlinecite{Mayers09:xx} calculates the energy widths by assuming that the Impulse approximation is valid.  This assumption is also made to obtain the energy profiles of a constant momentum scan using an indirect geometry spectrometers. If as found in Ref. \onlinecite{Mayers09:xx} the hydrogen cross section is not constant with momentum transfer, then the impulse approximation needs to be revaluated as done in several of the theories and papers discussed in the comment.  The result is that assuming the impulse approximation to analyse the data and then deriving a momentum dependent cross section is inconsistent and we believe this analysis needs to be reconsidered.   

	In contrast with a direct geometry spectrometer, we are able to measure the widths directly from constant angle scans without using the impulse approximation because the energy transfer of the scattering is centered at the free hydrogen energy.  

\subsection{2) Hydrogen cross section is constant as a function of both momentum and energy transfer}

	Our published paper shows firstly that the cross section for all momentum transfers and energy resolutions is constant.  Secondly, the absolute value of the cross section is that expected based on the Born and Impulse approximations.   We obtain the results independent of the incident neutron energy, independent of the energy resolution and independent of the scattering angle.  We note that Ref. \onlinecite{Mayers09:xx} agrees that the energy widths of the hydrogen recoil line are comparable on both spectrometers at high scattering angles, yet we derive the same cross section at these angles for a variety of incident neutron energies and momentum transfers.   We therefore do not agree with the comments in Ref. \onlinecite{Mayers09:xx}  as we have found that the hydrogen cross section is constant within the experimental error for all angles and momentum transfer and independent of the incident neutron energy.
	
\subsection{3) Jacobian}

	We do not agree with the semantics used by the authors when discussing resolution and believe it to be misleading.    When discussing resolution applied to neutron inelastic scattering, there are two key points- first the raw width of the resolution ellipsoid in momentum and energy, and second how the resolution ellipsoid cuts the dispersion surface of the excitation being measured.  The later point is defined by the Jacobian discussed in Ref 1.  We believe this second point is the origin of the broad peaks observed in a constant angle scan obtained with an indirect spectrometer in Ref. \onlinecite{Mayers09:xx}.

	Ref. \onlinecite{Mayers09:xx} present data based on simulated constant momentum scans.  We believe this analysis is invalid as the authors have assumed the impulse approximation to obtain these plots.    Also, the measurements deriving the change in the cross section were obtained from constant angle scans and not by the scans presented.

\subsection{4) Integrated intensity and sum rules}
	
	Ref. \onlinecite{Mayers09:xx} refer in their comment to several theories stating that the momentum cross section is tied to the energy resolution.  Such a statement is consistent with sum rules which state that the integral over energy must be a constant for different momentum transfers.  The data sets presented are not consistent with this sum rule because the integral of the scattered intensity is not independent of the momentum transfer.  The data taken on the direct instrument, MARI, is consistent with this basic notion of neutron scattering. 

	If the claim in Ref. \onlinecite{Mayers09:xx} that the apparent inconsistency between the results of direct and indirect geometry machines is due to the different energy resolutions of the experiments, then by integrating over all energies (or time) they should be able to obtain where the missing intensity has reappeared.  This analysis has never been performed to our knowledge.
	
	Theoretical work in Ref. \onlinecite{Karlsson11:xx} provided a possible explanation for some of the measurements described in Ref. \onlinecite{Mayers09:xx}.  However, the suggestions in Ref. \onlinecite{Karlsson11:xx} are inconsistent with our measurements because we observe the same intensity for different incident neutron energies and for a range of scattering angles. If the theory in Ref. \onlinecite{Karlsson11:xx} was an appropriate description, the changes in the intensity would be observed in our experiment and the missing intensity could be derived from the data taken on Vesuvio.

\subsection{5) Inconsistent Experimental results in the literature}

	A survey of the literature shows that the hydrogen cross section has been measured using the instrument VESUVIO at different times over the past 12 years and the missing intensity has tended to become smaller.  We consider that this suggests that the results are probably caused by experimental issues, rather than physical ones such as the method of analyzing the data.  This point can be explicitly seen in comparing the results of Refs. \onlinecite{Vos05:227} and \onlinecite{Cowley06:18}   which report very different values and angular dependences of the hydrogen cross section in polyethylene.    The results are also very different from the reports of experiments on various hydride materials.~\cite{Redah05:790}  This result again seems inconsistent with the Impulse approximation where all the atoms should respond independently at short times and therefore all hydrogen containing materials should give consistent results.  We believe the authors need to give the experimental setup, particularly given the recent upgrade outlined in a recent publication.~\cite{Mayers11:22}  Only when all of the data (widths and intensities) are provided can we determine if there is a discrepancy between experiments and theory in the current experimental configuration.
	
	Based on these five points, we do not agree with the comments described in Ref. \onlinecite{Mayers09:xx} and consider that the conclusion that the cross section of hydrogen varies with momentum transfer to be an experimental artifact associated with indirect geometry spectrometers.

%\end{thebibliography}

\end{document}